\documentclass[aps,prb,twocolumn,showpacs]{revtex4}

\usepackage{graphicx}

\def\bra{\langle}
\def\ket{\rangle}

\def\vG{{\bf G}}

\def\vI{{\bf I}}

\def\vV{{\bf V}}

\def\vg{{\bf g}}

\def\vk{{\bf k}}

\def\vr{{\bf r}}

\begin{document}

\title{Van Hove features in Bi$_2$Sr$_2$CaCu$_2$O$_{8+\delta}$ and
 effective parameters for Ni impurities inferred from STM spectra}

\author{Jian-Ming Tang and Michael E. Flatt\'e}

\affiliation{Department of Physics and Astronomy, University of Iowa,
  Iowa City, IA 52242-1479}

\date{submitted to PRB on 3 May 2002}

\begin{abstract}
  We present a detailed quantitative comparison between theoretical
  calculations of the local density of states and recent experimental
  measurements of scanning tunneling spectra around Ni impurities in
  Bi$_2$Sr$_2$CaCu$_2$O$_{8+\delta}$. A double-peak structure on the
  hole side of the spectrum at the Ni site is identified as the
  spin-split van Hove singularity in the band structure. The Ni atom
  induces local changes in hopping matrix elements comparable in size
  to the induced on-site spin-dependent potential. We find evidence
  from the measurements of order parameter suppression in the vicinity
  of the Ni impurity. These extracted impurity parameters can be of
  use in quantitative calculations of macroscopic response properties,
  such as the AC conductivity.
\end{abstract}


\maketitle

The response of a superconductor to impurities can determine several
macroscopic electromagnetic properties, such as the magnetic
penetration depth and high frequency losses.\cite{Tinkham1995} For
high-temperature superconductors it is advantageous to study the
impurity problem in real space rather than in momentum
space,\cite{Flatte1999} not only because the impurities destroy the
translational invariance, but also because these doped systems are
intrinsically inhomogeneous.\cite{Pan2001} High-quality electronic
spectra, using the technique of scanning tunneling microscopy (STM),
around individual impurity atoms in superconducting
Bi$_2$Sr$_2$CaCu$_2$O$_{8+\delta}$ have recently become
available.\cite{Pan2000,Hudson2001} Quantitative comparison between
calculations of the local density of states (LDOS) and these STM
measurements can lead to effective impurity parameters that are useful
for calculating macroscopic properties.

A number of theoretical studies of the STM spectra near impurities in
$d$-wave superconductors have already been carried out by various
authors.\cite{Byers1993,Salkola1996,Flatte1998,Flatte1999,Haas2000,Flatte2000,Polkovnikov2001,Zhu2001}
These previous studies focused on qualitative features, such as the
existence of quasiparticle resonances, the spatial structure of the
resonances, and the differences among systems with various pairing
symmetries. Often the model for an impurity only consisted of
effective potentials at the impurity site, or relied on an unrealistic
band structure for the host superconductor. There are still
significant quantitative disagreements between the theoretical and
experimental LDOS.\cite{Flatte2000,Polkovnikov2001,Zhu2001} Of the
types of impurities and defects for which high-quality STM spectra are
available, the Ni impurity provides a good opportunity to resolve some
of these issues. Because the perturbation to the local environment
caused by a Ni impurity is weak,\cite{Hudson2001,Flatte2001} we find
that, with some extensions to the potential model the STM spectra can
be well reproduced by our calculations. In addition to the
short-ranged potentials at the impurity site, we also include the
change of the hopping matrix elements from the impurity site to its
nearest neighbors, and the suppression of the order parameter.

We also find an interesting interplay between the underlying band
structure and the superconducting coherence peak near a Ni
impurity. In the spectrum of the Ni site there is a large peak on the
hole side (below the chemical potential), right outside the gap
edge. We find that this peak, previously thought as the remnant of the
coherence peak, is largely contributed by a density-of-states peak in
the underlying band structure. One special and important feature of
hole doped cuprate superconductors is that a van Hove singularity is
close to the Fermi level. It has been suggested by various
authors\cite{Hirsch1986,Lee1987,Labbe1987,Friedel1989,Markiewicz1990,Getino1992,Newns1992}
that the existence of the van Hove singularity could affect the
mechanism of superconductivity, and also account for the unusual
normal state properties. The existence of such a density-of-states
peak has received direct support from experiments using angle-resolved
photoemission spectroscopy (ARPES).\cite{Shen1995,Ma1995} Despite this
density-of-states peak (aside from the coherence peaks) in the
superconducting state has never been directly reported in tunneling
experiments. Traces of this feature were suggested to explain the
asymmetry of the coherence peaks,\cite{Renner1995,Bok1997} and the
sharpness of the coherence peaks. However, no specific qualitative
influence of a van Hove peak on the midgap impurity resonances of a
particular impurity has been considered.\cite{Crisan1999}

We find that the Ni impurity offers a novel probe of the van Hove
peak. The van Hove singularity is likely very close to the Fermi
level, therefore, in the superconducting state the van Hove peak
cannot be resolved from the coherence peak. The Ni impurity will
locally suppress the coherence peak, but only weakly perturb the
underlying band structure. As a result, the van Hove peak shows up
unmasked in the spectrum at the impurity site.

\begin{figure}
\includegraphics[width=0.9\columnwidth]{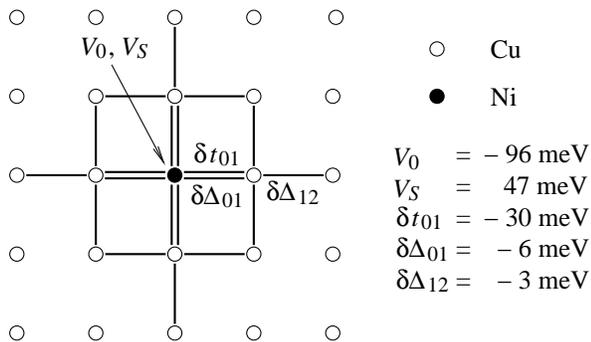}
\caption{ A schematic diagram showing our model of a Ni impurity on the
  Cu--O plane. Parameters extracted from fitting to the data are
  listed on the right. }
\label{fig:model}
\end{figure}

Our calculation is based on the following tight-binding Hamiltonian,
\begin{eqnarray}
H & = & -\sum_{\bra i,j\ket,\,\sigma}(t_{ij}+\delta t_{ij})c^\dagger_{i\sigma}c_{j\sigma} \nonumber\\
&& +\sum_{\bra i,j\ket}\left[(\Delta_{ij}+\delta\Delta_{ij})c^\dagger_{i\uparrow}c^\dagger_{j\downarrow}+(\Delta^*_{ij}+\delta\Delta^*_{ij})c_{j\downarrow}c_{i\uparrow}\right] \nonumber\\
&& +(V_{0}+V_{S})c^\dagger_{0\uparrow}c_{0\uparrow}+(V_{0}-V_{S})c^\dagger_{0\downarrow}c_{0\downarrow} \;,\label{eq:H}
\end{eqnarray}
where $i$ and $j$ label sites (the impurity resides at site 0), and
$\sigma$ labels spin. The hopping matrix elements, $t_{ij}$, are taken
from a one-band parameterization of the ARPES data.\cite{Norman1995}
The numerical values in units of meV are $148.8$, $-40.9$, $13$, $14$,
and $-12.8$ for the hopping to the first five nearest neighbors. The
measured value for the gap maximum $\Delta_0$ at a site away from the
impurity is $28$ meV.\cite{Hudson2001} The order parameters,
$\Delta_{ij}$, are assumed to be real, and to have $d$-wave
symmetry. They are only non-zero on the bonds between two
nearest-neighbor sites, and $\Delta_{i,i+\hat x}=-\Delta_{i,i+\hat
y}=\Delta_0/4$. The momentum-dependent order parameter resulting from
a homogeneous system with these $\Delta_{ij}$ is
$\Delta_\vk=(\Delta_0/2)(\cos k_x-\cos k_y)$, where the lattice
spacing between two Cu atoms is unity.

Because the experiments were carried out at low temperatures, the spin
degree-of-freedom of the magnetic Ni impurity is assumed to be
completely frozen. At the Ni site, we use a potential that consists of
a magnetic part, $V_S$, and a non-magnetic part, $V_0$. It is found
that the Ni impurity induces a change in the hopping, $\delta t_{ij}$,
to the nearest neighbor sites, and a suppression of the order
parameter magnitudes, $\delta\Delta_{ij}$, on the four bonds connected
to the Ni site, and on the other twelve bonds connected to the nearest
neighbor sites. A schematic diagram of our model and the fitting
parameters used for the Ni impurity are shown in FIG.~\ref{fig:model}.

Calculations presented here are based on a Koster-Slater
technique.\cite{Flatte1997a,Flatte1997b,Flatte1999} The homogeneous
Green's function, $\vg(\vr_i,\vr_j,\omega)$, in the Nambu formalism is
first constructed with a $1$ meV quasiparticle linewidth. In the
presence of an impurity modeled by $\vV$, the Gor'kov equation,
$\vG=\vg+\vg\vV\vG=(\vI-\vg\vV)^{-1}\vg$, is solved by numerically
inverting the matrices. The LDOS is the imaginary part of the Green's
function, $(-1/\pi){\rm
Im}\left[\vG_{11}(\vr_i,\vr_i,\omega)-\vG^*_{22}(\vr_i,\vr_i,-\omega)\right]$.

\begin{figure}
\includegraphics[width=\columnwidth]{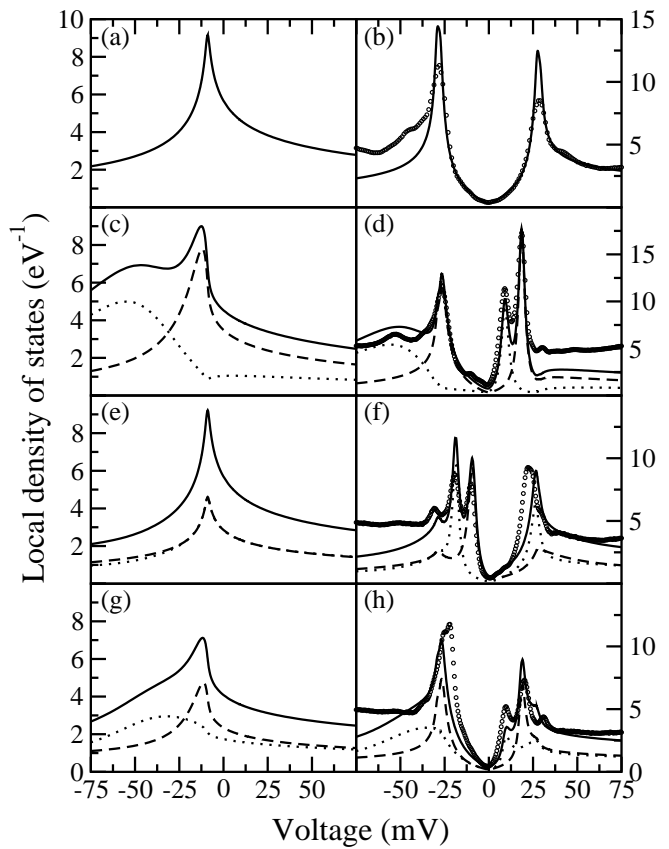}
\caption{ LDOS spectra at various sites near a Ni impurity. The left
  panels show the spectra in the normal state. The right panels show
  the spectra in the superconducting state. Solid lines show the
  calculated results. Dashed and dotted lines show the two spin
  components. (a) and (b) are the spectra at a site about $30$\AA\
  away from the Ni impurity. (c) and (d) are the spectra right at the
  impurity site. (e) and (f) are the spectra on the nearest neighbor
  site. (g) and (h) are the spectra on the second nearest neighbor
  site. Open circles ($\circ$) show the STM data.\cite{Hudson2001} The
  data was rescaled by a constant factor identical for all the
  spectra. }
\label{fig:sites}
\end{figure}

\begin{figure}
\includegraphics[width=\columnwidth]{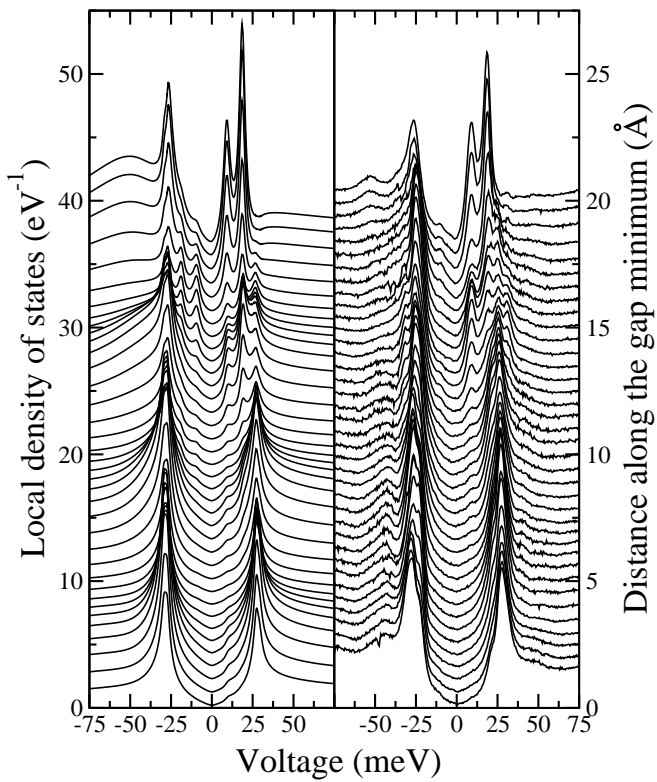}
\caption{ Evolution of the spectra as a function of distance from the
  Ni atom. Left panel shows the calculations, and the right panel
  shows the data from Fig.~4 in Ref.~5. The spin splitting of the
  resonance peaks inside the gap is constant as one moves away from
  the impurity site. The spin splitting of the van Hove singularity
  decreases as the distance increases.  }
\label{fig:lines}
\end{figure}

In the following, we will discuss how various parameters influence our
calculations. Although six parameters including the chemical potential
are introduced in the model, there are particular features in the LDOS
spectra associated with each of them. This allows us to determine each
parameter largely independently from the others.

Let us first consider the chemical potential. From the ARPES
parameterization,\cite{Norman1995} the chemical potential for an
optimally doped sample is $-130.5$ meV. However, we know from STM
measurements that the doping concentration is locally varying, and is
correlated with the superconducting gap magnitude.\cite{Pan2001} A
region with a gap magnitude of $28$ meV is overdoped. The fit to STM
spectra is optimal if the shift in chemical potential relative to the
optimally doped region is about $-25$ meV. This amount of shift
changes the location of the van Hove singularity from $34$ meV to $9$
meV below the Fermi level, shown in FIG~\ref{fig:sites}(a) for
$\Delta_{ij}=0$.\cite{NormalState} In the superconducting state the
van Hove singularity cannot be resolved from the coherence peak at the
hole side, shown in FIG~\ref{fig:sites}(b). The chemical potential
shift also sharpens the coherence peaks, and contributes to the
asymmetry between the two peaks. Although not direct evidence for the
van Hove singularity in STM, the agreement between the calculations
and the measurements is certainly improved.

We now describe the modeling of the Ni impurity focusing first on the
single-site potential model with only the on-site potentials, $V_0$
and $V_S$. For a weak potential, the main effect of these potential is
to set the energies of the resonances. Since the resonance peaks for
Ni are at $9.2$ and $18.6$ meV, and the perturbation from Ni is weak,
both $V_0+V_S$ and $V_0-V_S$ should be negative. Our $V_0$ does not
differ greatly from that found in a simple
model.\cite{Fehrenbacher1996} The two on-site parameters, $V_0$ and
$V_S$, are both required to fit the energies of the two resonance
peaks.\cite{Salkola1997} Note that we have no control over the peak
height or width in the spectra using this simple model. We find, due
to extended states within the gap, that the resonance peaks will be
broader than the coherence peaks. In the observation, however, the
width of the resonance peaks appears comparable, or even sharper, than
the coherence peaks.

In order to reduce the coupling to the extended states and increase
the localization of the resonance states, we then assume that the
hopping to the nearest neighbor sites, $\delta t_{01}$, is reduced.
With only a $20\%$ change in the nearest-neighbor hopping, we can
reproduce the linewidth of the resonance peaks. $V_0$ and $V_S$ must
be adjusted slightly when $\delta t_{01}$ is introduced in order to
keep the resonances at the fixed energies. Our calculations show that
the magnitudes of the on-site potentials change to even weaker values,
and characterize an even weaker impurity potential. Using these three
parameters for calculating the effect of a Ni impurity in the normal
state (shown in FIG~\ref{fig:sites}(c)), one can see that the band
structure is not highly distorted. The van Hove peak remains
reasonably sharp, and splits into two spin components. Comparing
\ref{fig:sites}(c) to the superconducting state, shown in
FIG~\ref{fig:sites}(d), we can identify the two peaks on the hole
side, right outside the gap edge, as the spin-split van Hove peaks.  A
similar connection between the normal state electronic structure and
an impurity's influence on the superconducting state has been
identified in calculations for impurities in s-wave
superconductors.\cite{Flatte1997a,Flatte1997b} Another distinction
between the midgap resonances and the van Hove peaks is that the spin
splitting of the resonance peaks inside the gap remains independent of
distance from the impurity site. The spin splitting of the van Hove
singularity, however, decreases as the distance increases.

Once the large peak at the hole side was identified as the van Hove
peak, a noticeable disagreement between the calculation and the data
became apparent. The calculated relative energy difference between the
resonance peaks and the van Hove peaks is somewhat larger than the
observed value if the gap maximum is constant everywhere. In the
$d$-wave superconducting state, the position of the van Hove peak is
dominated by the gap edge. One cannot increase the agreement by tuning
$V_0$ because that shifts all peaks almost an equal amount of
energy. We could obtain good agreement if the gap maximum were about
$22$ meV instead of $28$ meV. In principle using self-consistent
calculations can account for the effect, however, the self-consistency
conditions involve the pairing mechanism, which is
controversial. Taking a phenomenological approach, we can identify the
suppression of the order parameter around the impurity from the
data. This then becomes a constraint on pairing theory. We find that
the parameter is locally suppressed more than $50\%$ (See
FIG.~\ref{fig:model}).

\begin{figure}
\includegraphics[width=\columnwidth]{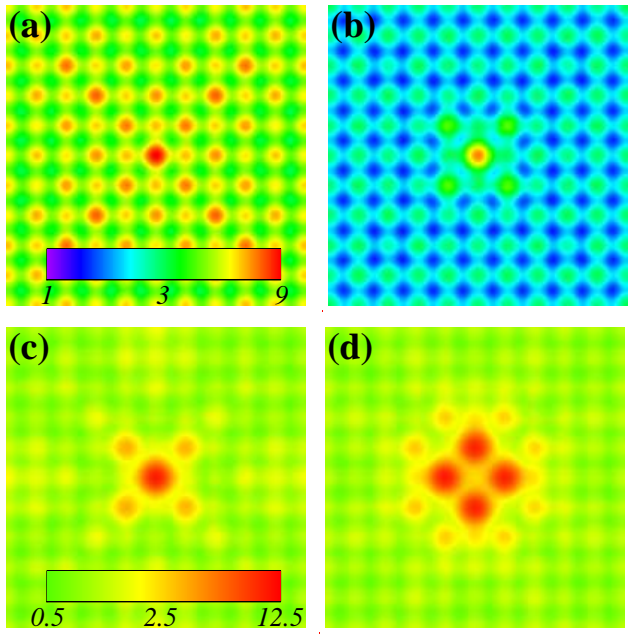}
\caption{ (color) Spatial structure of the LDOS around a Ni impurity
  in the normal state at (a) $-15$ meV, (b) $-50$ meV (near van Hove peak
  positions in FIG.~\ref{fig:sites}(c)), and in the superconducting
  state at (c) $+9$ meV, and (d) $-9$ meV. The Wannier function is
  assumed to be a Gaussian with a width of half the lattice
  spacing. The LDOS is shown in logarithmic scales and in units of
  eV$^{-1}$. The lattice orientation in (c) and (d) needs to be
  rotated by $45^\circ$ when compared to Fig.~2 in Ref.~5. }
\label{fig:maps}
\end{figure}

So far, we were only fitting the spectrum right at the Ni site.
Nevertheless, good agreement in spatial structure is also obtained.
Detailed comparisons for the spectra on neighboring sites are shown in
FIG.~\ref{fig:sites}(e)--(h), the spatial decay of the resonance peaks
is shown in FIG.~\ref{fig:lines}, and the spatial maps at given
energies are shown in FIG.~\ref{fig:maps}. These are direct
comparisons with the data in FIGs.~2--4 in Ref.~5.

Two key additional features are found from the spatial structure. The
first one is that the hole component of the resonance has $d$-wave
symmetry.\cite{Flatte1999,Flatte2001} The second is that the
perturbation, due to the van Hove singularity, is strongest along the
diagonals of the lattice. At given energy, the secondary peaks are
always located at the next-nearest rather than the nearest neighbor
sites relative to the main peaks. This is because, around the Fermi
level, the saddle points that give rise to the van Hove singularity in
the density of states are near the $(\pm\pi,0)$ and $(0,\pm\pi)$
points in momentum space. The main contribution to the real space
Green's function largely comes from these high density-of-states
regions,
\begin{equation}
\vg(\vr_i,\vr_j,\omega) = \int \frac{d^2\vk}{(2\pi)^2}e^{i\vk\cdot(\vr_i-\vr_j)}\vg(\vk,\omega) \;.
\end{equation}
For $\vr_i-\vr_j=(\pm 1,\pm 1)$ the four regions add up
constructively, while for $(\pm 1,0)$ or $(0,\pm 1)$, the
contributions from $(\pm\pi,0)$ cancel with those from
$(0,\pm\pi)$. As a result, $|\vg((1,1),(0,0),\omega)|\gg|\vg((1,0),(0,0),\omega)|$ for
energies near the chemical potential. This is also consistent with the
larger spin splitting of the van Hove peak at the second nearest
neighbors compared to the nearest neighbors.

In this study we have assumed that the STM spectra are directly
related to the electronic structure of the Cu--O plane regardless of
the presence of the intermediate Bi--O plane. As was argued in Ref.~5,
the energies and linewidths of the resonances should be the same. We
find the principal effect of a $d$-wave filter\cite{Martin2002} on our
parameterization is to (roughly) change the sign of $V_0$. Therefore,
the evidence for the reduction of hopping, and suppression of the
order parameter are still clear.

In summary, the presence of a Ni atom on the Cu--O plane produces a
weak perturbation to the system. We show direct evidence for a
density-of-states peak in the STM measurements and local suppression
of the order parameter. The suppression of the order parameter can
constrain the self-consistency condition and thus the pairing
mechanism.

We thank E. W. Hudson and J. C. Davis for providing data shown in
FIGs.~\ref{fig:sites}--\ref{fig:lines} from Ref.~5.  This work is
supported by ONR Grant No.~N00014-99-0313.


\begin{thebibliography}{32}
\bibitem[{Nor()}]{NormalState}
\bibinfo{note}{The normal state in cuprates is known to have a variety of
  unusual properties. Thus the state with $\Delta_{ij}=0$ may or may not
  correspond to the true normal state of the solid. We will, however, use the
  term ``normal state'' in this paper to refer to that state with the
  Hamiltonian of Eq.~(\ref{eq:H}), but with $\Delta_{ij}=0$.}
\end{thebibliography}
\end{document}